\documentclass[twocolumn,prl,showpacs,superscriptaddress]{revtex4}

\usepackage{graphicx}
\usepackage{amssymb}
\usepackage{amsmath}

\begin{document}

\title{Aging dynamics in interacting many-body systems}

\author{Lloyd P. Sanders}
\affiliation{Department of Astronomy and Theoretical Physics, Lund University,
Lund, Sweden.}
\author{Michael A. Lomholt}
\affiliation{MEMPHYS - Center for Biomembrane Physics, Department of Physics
and Chemistry, University of Southern Denmark, Odense, Denmark}
\author{Ludvig Lizana}
\affiliation{Integrated Science Lab, Department of Physics, Ume{\aa}
University, Ume{\aa}, Sweden}
\author{Karl Fogelmark}
\affiliation{Department of Astronomy and Theoretical Physics, Lund University,
Lund, Sweden.}
\author{Ralf Metzler}
\affiliation{Institute for Physics \& Astronomy, University of Potsdam,
Potsdam-Golm, Germany}
\affiliation{Department of Physics, Tampere University of Technology,
Tampere, Finland}
\author{Tobias Ambj\"ornsson}
\email{tobias.ambjornsson@thep.lu.se}
\affiliation{Department of Astronomy and Theoretical Physics, Lund University,
Lund, Sweden.}

\date{\today}

\begin{abstract}
  Low-dimensional, complex systems are often characterized by
  logarithmically slow dynamics. We study the generic motion of a
  labeled particle in an ensemble of identical diffusing particles
  with hardcore interactions in a strongly disordered, one-dimensional
  environment. Each particle in this single file is trapped for a
  random waiting time $\tau$ with power law distribution
  $\psi(\tau)\simeq\tau^{-1- \alpha}$, such that the $\tau$ values are
  independent, local quantities for all particles. From scaling
  arguments and simulations, we find that for the scale-free waiting
  time case $0<\alpha<1$, the tracer particle dynamics is ultra-slow
  with a logarithmic mean square displacement (MSD) $\langle
  x^2(t)\rangle\simeq(\log t)^{1/2}$.  This extreme slowing down
  compared to regular single file motion $\langle x^2(t)\rangle\simeq
  t^{1/2}$ is due to the high likelihood that the labeled particle
  keeps encountering strongly immobilized neighbors. For the case
  $1<\alpha<2$ we observe the MSD scaling $\langle x^2(t)\rangle\simeq
  t^{ \gamma}$, where $\gamma<1/2$, while for $\alpha>2$ we recover 
  Harris law $\simeq t^{1/2}$.
\end{abstract}


\maketitle

Ultraslow, logarithmic time evolution of physical observables is remarkably
often observed, for instance, for paper crumpling in a piston \cite{paper},
DNA local structure relaxation \cite{dna}, frictional strength
\cite{friction}, grain compactification \cite{grain}, glassy systems
\cite{glasses}, record statistics \cite{record}, as well as magnetization,
conductance, and current relaxations in superconductors, spin glasses, and
field-effect transistors \cite{magnetization}.  Here we demonstrate how
ultraslow dynamics of a labeled particle in a many-body system of excluded
volume particles arises while without the excluded volume effects the dynamics
is characterized by a power-law spreading.

Imagine a single colloidal particle diffusing in a narrow fluidic
channel. Its mean squared displacement (MSD) $\langle x^2(t)
\rangle\simeq t$ will display the linear time dependence
characteristic of Brownian motion. In contrast, if the same particle
is made viscid by functionalization with ``sticky'' ends and the channel surface
complementary coated, its motion will exhibit intermittent pausing
events caused by transient binding to the channel surface. The
distribution of pausing durations $\tau$ is of power-law form $\psi
(\tau)\simeq\tau^{-1-\alpha}$ with $0<\alpha<1$ in a certain
temperature window, effecting subdiffusive behavior $\langle
x^2(t)\rangle\simeq t^{\alpha}$ \cite{xu}.  This random motion belongs
to the family of the Scher-Montroll-Weiss continuous time random walk
(CTRW) \cite{montroll,scher}, a renewal process with independent
successive waiting times.
\begin{figure}
\includegraphics[width=8cm]{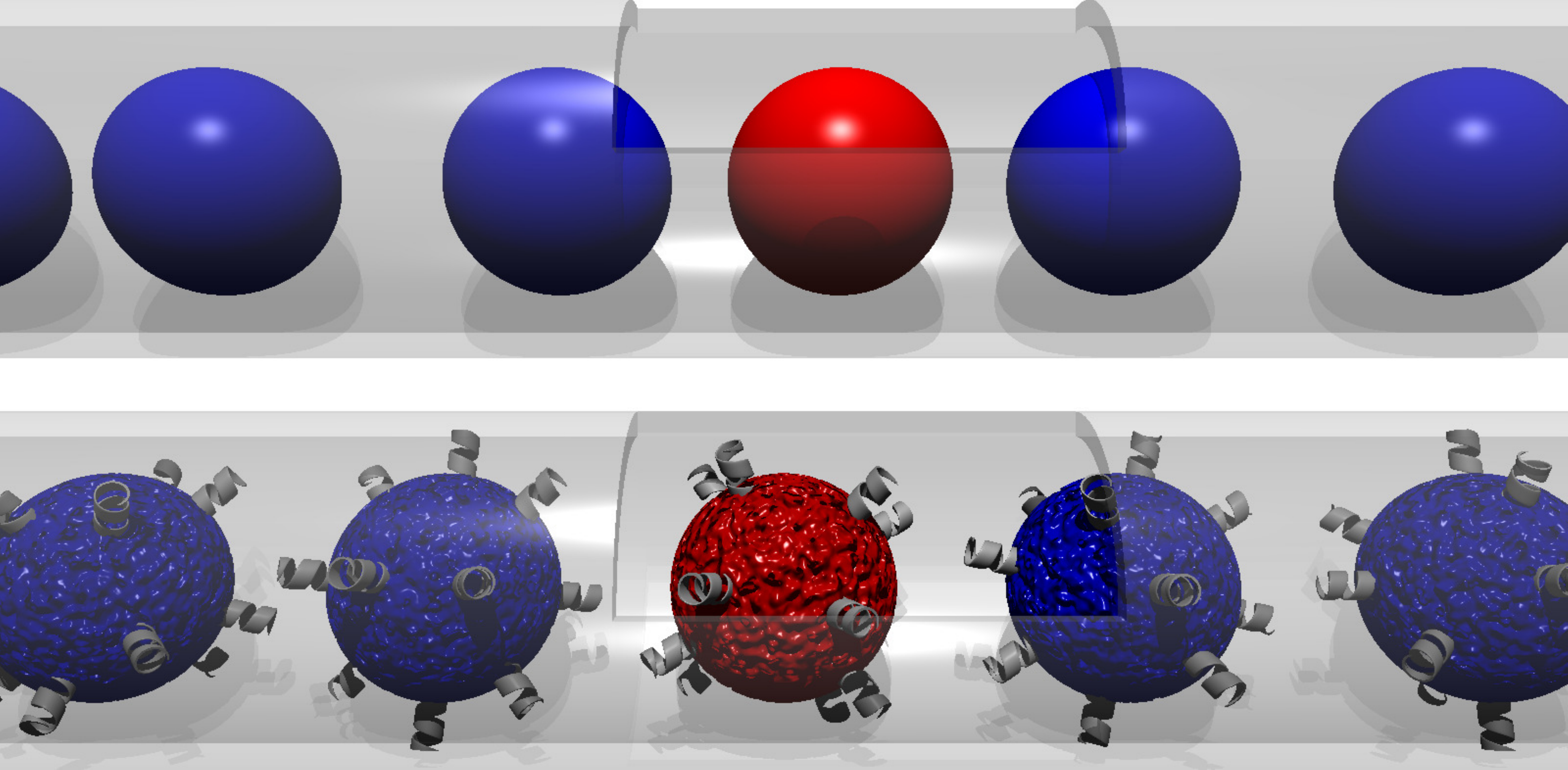}
\caption{Schematic of a narrow channel containing a single file of $N$ colloidal
particles particles. Regular particles in a bare channel perform single file
motion characterized by Harris' law $\langle x^2(t)\rangle\simeq t^{1/2}$ (top),
in the case of functionalized, sticky particles (bottom) the motion becomes
ultraslow, $\langle x^2\rangle\simeq\log^{1/2}(t)$.}
\label{fig:CTRW}
\end{figure}
Stochastic dynamics governed by power-law forms of $\psi(\tau)$, with
$0<\alpha<1$ were also shown to apply to tracer particle motion in the
cytoplasm \cite{lene} in membranes \cite{krapf} of living cells, in
reconstituted actin networks \cite{weitz}, and determine the blinking
dynamics of single quantum dots \cite{qdots} as well as the dynamics
involved in laser cooling \cite{aspect}. Physically, the form
$\psi(\tau)$ may arise from comb models \cite{havlin} or random energy
landscapes \cite{rel}. The divergence of the mean waiting time
$\langle\tau\rangle=\int_0^{\infty}\tau\psi( \tau)d\tau$
\cite{report,hughes} leads to ageing phenomena \cite{ageing} and weak
ergodicity breaking \cite{web}, with profound consequences for, e.g.,
molecular cellular processes \cite{pt}.

What will happen if we surround the colloidal particle with identical
particles (Fig.~\ref{fig:CTRW})?  As the channel is narrow, individual
particles cannot pass each other, thus forming a single file
\cite{harris,levitt,rodenbeck,lizana}. When the colloidal particles and the
channel walls are not coated, it is well known that the mutual exclusion of
the particles in the channel leads to the characteristic Harris scaling $\langle
x^2(t)\rangle\simeq t^{1/2}$ of the MSD of the labeled particle
\cite{harris,levitt,rodenbeck,lizana}. However, once we introduce sticky
surfaces,  we will show
from scaling arguments and extensive simulations that the motion of the
labeled, sticky particle becomes slowed down dramatically, its MSD following
the logarithmic law $\langle x^2(t)\rangle\simeq\log^{1/2}(t)$. We
also find that even when a characteristic waiting time $\langle\tau\rangle$
exists, as long as $1<\alpha<2$ the motion of the labeled particle is still
anomalous, with a dynamic exponent $\gamma<1/2$. Only when $\alpha>2$, we
return to the regular $1/2$ Harris scaling exponent.

\emph{Motion rules for the walkers.} A \emph{single walker\/} is updated
following the simple CTRW rules: on our one-dimensional lattice, jumps occur
to left and right with equal probability, and the waiting times between
successive jumps are drawn from the probability density
\begin{equation}
\label{wtd}
\psi(\tau)=\alpha/[\tau^{\star}(1+\tau/\tau^{\star})^{1+\alpha}],
\end{equation}
where $\tau^{\star}$ is a scaling factor with unit of
time. Practically, the waiting times become
$\tau=\tau^{\star}[r^{-1/\alpha}-1]$, where $r$ is a uniform random
number from the unit interval. After each jump the walker's clock is
updated, algorithmically, $T\rightarrow T+\tau$, where initially
$T=0$. For the scale-free case, $0<\alpha<1$, i.e., infinite average
waiting time $\langle\tau\rangle$, we obtain subdiffusive transport,
$\langle x^2(t)\rangle=2K_{\alpha}t^{\alpha}/ \Gamma(1+ \alpha)$
\cite{report}. Here,
$K_{\alpha}=a^2/[2(\tau^{\star})^{\alpha}\Gamma(1- \alpha)]$ is the
anomalous diffusion coefficient with $a$ the lattice spacing. For
$\alpha>1$, $\langle\tau \rangle$ is finite and we recover normal
diffusion.

\begin{figure}
\includegraphics[width=8cm]{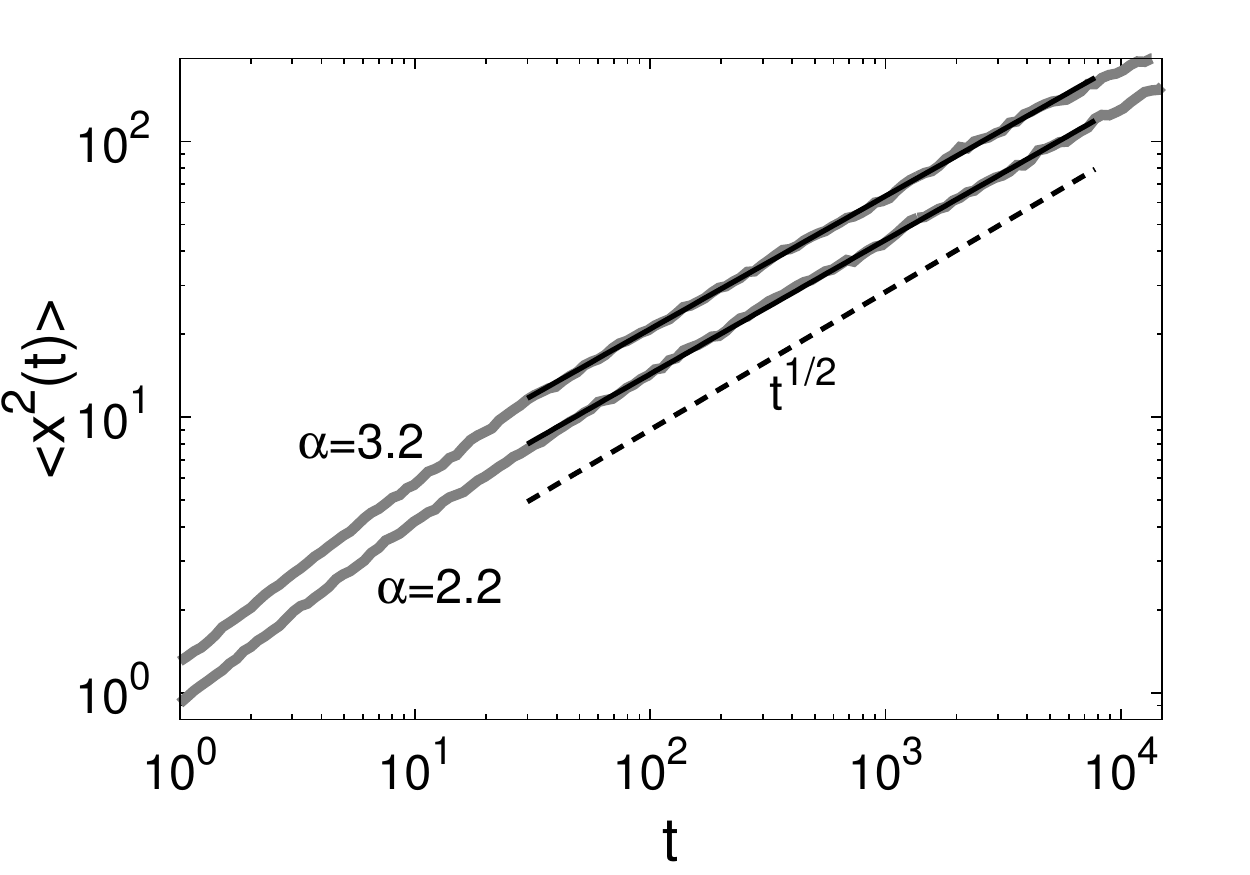}
\caption{MSD $\langle x^2(t)\rangle$ for a single file system with waiting
  time distribution (\ref{wtd}), and $\alpha=$ $2.2$ and $3.2$. In both cases
  the MSD follows the Harris $1/2$-scaling (dashed line) for long
  times. Parameters: scaling factor $\tau^{\star}=1$, lattice size $L=600$,
  number of particles $N=201$, so that the particle density is
  $\varrho=N/L\approx 0.3$. The MSD was averaged from $3.5\times10^3$
  simulations.}
\label{fig:plot_LT}
\end{figure}

For the case of a single file of {\em many} excluded-volume walkers the motion
of individual particles is updated in a similar fashion, with the convention
that any jump leading to a double occupancy of lattice sites is canceled.
More specifically: (i) Assign initial positions to
all the $N$ particles (indexed by $i=1,...,N$). We position the labeled particle
at the middle lattice point and randomly distribute equally many particles to
the left and right. Each particle carries its own clock with timer $T_i$, and
all clocks are initiated simultaneously, $T_i=0$.  (ii) Draw an independent
random waiting $\tau_i$ from Eq. ({\ref{wtd}) for
  each particle and add this to the timer, $T_i\to T_i+\tau_i$.  (iii)
  Determine the particle $j$ with the minimal value $T_j=\min\{T_i\}$ and move
  particle $j$ with probability $1/2$ to the left or right, unless the chosen
  site is already occupied by another particle. In this case cancel the move.
  (iv) Add a new waiting time $\tau_j$ chosen from $\psi(\tau)$ to the timer
  of particle $j$, i.e., $T_j\rightarrow T_j+\tau_j$, and return to
  (iii). This is repeated until a designated stop time.

  This motion scenario used in our stochastic simulations directly
  reflects the local nature of the physical problem
  (Fig. \ref{fig:CTRW}). Namely, when we follow individual, sticky
  particles in the channel, each binding and subsequent unbinding
  event will provide a different, random, waiting time.  Even when the
  same position is revisited by the same particle, the waiting time
  will in general be different. To move the particle with the shortest
  remaining waiting time, i.e., whose timer first coincides with the
  laboratory (master) clock appears as a natural choice.

  For the case of the power-law waiting time distribution (\ref{wtd})
  with exponent $\alpha>2$, we show results from extensive simulations
  based on above motion rules in Fig.~\ref{fig:plot_LT}. Our results 
  reproduce the classical Brownian single-file scaling $\langle
  x^2(t)\rangle\simeq t^{1/2}$ \cite{harris}; the fitted scaling
  exponents are $0.49\pm0.01$ for $\alpha=2.2$ and $0.48\pm0.01$ for
  $\alpha=3.2$ \cite{finitesize}.

\begin{figure}
\includegraphics[width=8cm]{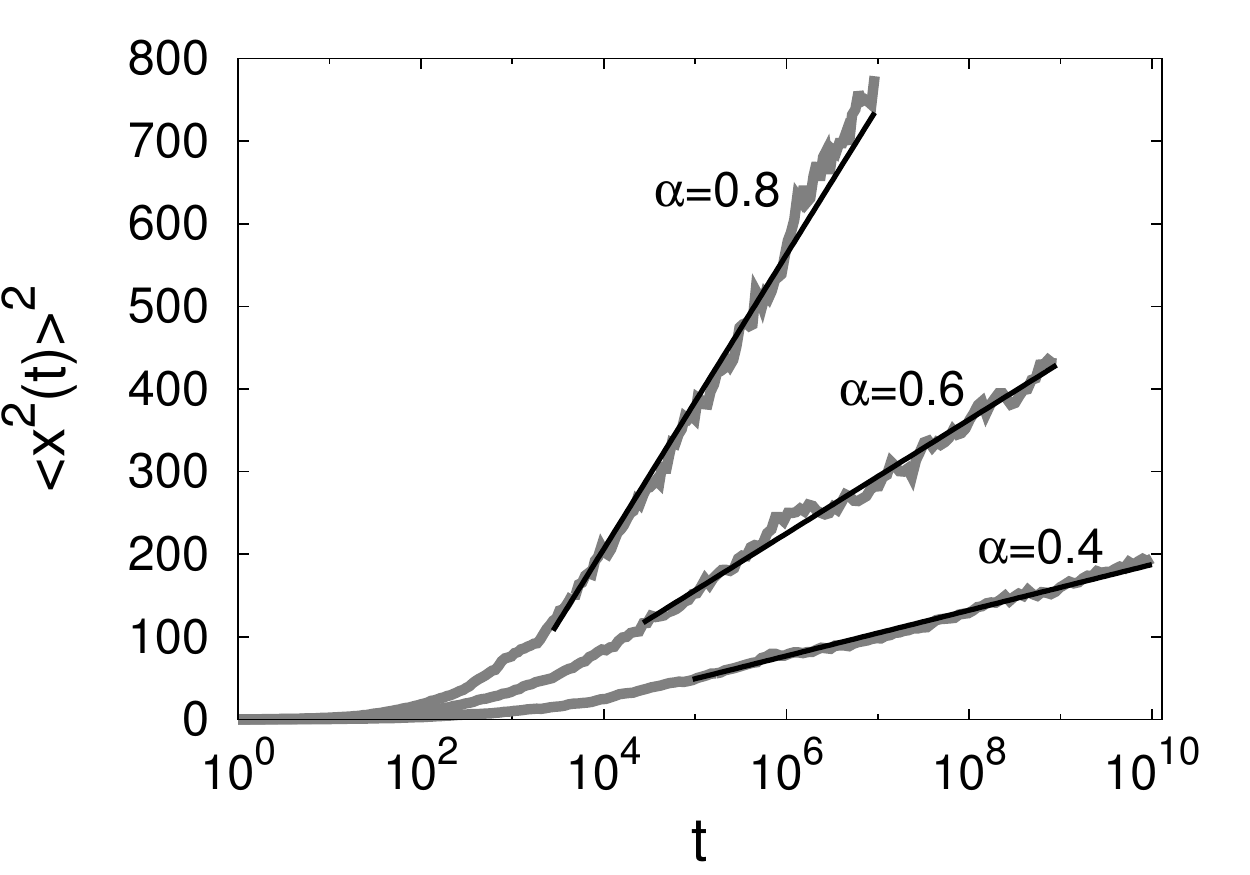}
\caption{Squared MSD (gray) for a labeled particle in a single file
  governed by the waiting time distribution (\ref{wtd}) with
  $0<\alpha<1$. Note the logarithmic abscissa. Black solid lines are
  fitted to $\langle x^2(t) \rangle=c_1\sqrt{\log t+c_2}$, see
  Eq.~(\ref{eq:MSD_HT}). Parameters: ensemble size $5\cdot 10^3$,
  lattice size $100$, number of particles $N=31$.}
\label{fig:plot_HT}
\end{figure}

Simulation results for $0<\alpha <1$ (diverging mean waiting time
$\langle\tau\rangle$) are depicted in Fig.~\ref{fig:plot_HT}. Note that we plot
the square of the MSD versus time, such that according to the logarithmic time
evolution, Eq. (\ref{eq:MSD_HT}), of the MSD derived from scaling argument below, we
would expect a linear dependence on the abscissa using linear-log
scales. Indeed, the numerical results strongly support the predicted universal
square root-logarithmic time evolution at long times for all exponents
$\alpha$.

To understand and quantify the system's dynamics we now obtain the MSD
for the labelled particle from a scaling argument. Let us start with
the case of a Poissonian (exponential) waiting time distribution with
a well-defined characteristic waiting time and finite moments of all
order. This is the scenario of regular single file diffusion with the
famed Harris' law \cite{harris}. If we consider the lattice dynamics
of the single file \cite{arratia}, the MSD of the labeled particle
reads
\begin{equation}
\label{eq:MSD_n}
\langle x^2(n)\rangle =Cn^{1/2},
\end{equation}
where we follow the evolution of the MSD as function of the number $n$ of
steps performed by the tracer particle. Given the finite mean waiting time
$\langle \tau\rangle$, the average number of steps is linearly related to the
process time $t$ through $\langle n\rangle=t/\langle\tau\rangle$. In
Eq.~(\ref{eq:MSD_n}) the prefactor is $C=a^2
(2/\pi)^{1/2}(1-\varrho)/\varrho$, where $\varrho$ is the particle
concentration (average number of particles per site).

Consider now the case of functionalized particles and channel walls
(non-Poissonian single file dynamics).
The dynamics is then characterized by motility periods, i.e. unhindered CTRW
motion described by Eq. (\ref{eq:MSD_n}), separated by blockage events
when immobile neighbors are encountered. For scale-free waiting time
distributions these blockage events are long compared with the duration of the
motility periods, and thus the blockage events dominate the dynamics. We take
this into account by converting the number of steps, $n$, in motility periods
into the process time, $t$, measured by the laboratory master clock
(subordination \cite{feller2}). Formally,
if we denote by $H_n(t)$ the probability that the tracer particle has taken
$n$ steps up to time $t$, we invoke the transformation
\begin{equation}
\label{eq:subordination}
\langle x^2(t)\rangle=\sum_n\langle x^2(n)\rangle H_n(t)\longrightarrow C\int_0^{
\infty}n^{1/2}H_n(t)dn.
\end{equation}
Here we assumed that the \emph{subordinator\/} $H_n(t)$ is slowly varying
in $n$ to replace the sum by an integral.

To proceed we employ a scaling argument to relate the number of steps $n$ with
laboratory time $t$ in the limit of many jumps (long times).  For a scale-free
distribution $\psi(\tau)$ of waiting times ($0<\alpha<1$), longer and longer
 $\tau$ occur in the course of the process. In particular,
individual $\tau$ may become of the order of the laboratory time. Thus, when
the labeled particle meets a trapped neighbor, statistically the neighbor will
experience one of these extremely long waiting time periods. Compared to these
extreme blockage events the local motion of the tracer particle shuttling back
and forth between immobilized neighbors will be negligible in the long time
limit. The duration of the limiting steps for the motion of the labeled
particle, i.e., to see its blocking neighbor resume its motion, are dominated
by the probability for the next jump to occur. This exactly corresponds to the
so-called forward waiting time of CTRWs \cite{ageing}. In the long time limit
we thus face a process, in which \emph{every\/} step $n$ is governed by the
forward waiting time. Such a process was considered recently, and there it was
shown that the average number of steps taken at time $t$ scales as $\log t$
\cite{lomholt}. Furthermore, the spread of the corresponding probability
distribution was shown to grow slower than $\log t$, implying that in the long
time limit we may consider $n\simeq\log t$ as a deterministic (scaling)
relation between $n$ and $t$.

The argument above leads us to the scaling ansatz for the subordinator
$H_n(t)$ for $0<\alpha<1$, namely, it should be expressed in terms of a
scaling function $f(n/\log t)$. Imposing the normalization
$\int_0^{\infty}H_n(t)dn=1$, i.e., a jump necessarily occurs at some given
time, we thus have the result $H_n(t)\simeq(\log t)^{-1}f(n/\log t)$ valid
in the limit of many jumps, $n\gg1$, which automatically implies $t\gg\tau^{
  \star}$. Combining this scaling form with Eqs.~(\ref{eq:MSD_n}) and
(\ref{eq:subordination}) we obtain, after a change of variables $n\rightarrow
n/\log t$, that
\begin{equation}
\label{eq:MSD_HT}
\langle x^2(t)\rangle\simeq[\log(t)]^{1/2}.
\end{equation}
Interestingly, compared to the standard square root scaling of Brownian single
file motion, the scale-free waiting time process introduces a logarithmic time.
As we show in Fig.~\ref{fig:plot_HT} this simple scaling argument combined with
the results from Ref.~\cite{lomholt} indeed accurately captures the dynamics
of the many-body CTRW system. We discuss this result further below.

What happens when we turn to larger values of the anomalous exponent $\alpha$,
such that the characteristic waiting time $\langle\tau\rangle$ becomes finite?
Similar to the observations in Ref.~\cite{lomholt} it turns out that we need to
distinguish two cases. Let us start with the case $\alpha>2$. The results of
\cite{lomholt} suggest a deterministic, linear scaling between $n$ and $t$.
Thus $H_n(t)\simeq t^{-1}f(n/t)$, i.e., Eq.~(\ref{eq:MSD_n}) becomes
\begin{equation}\label{eq:MSD_final}
\langle x^2(t)\rangle\simeq
t^{1/2}.
\end{equation}
and we recover the Brownian single file dynamics. This characteristic
$1/2$-scaling is indeed confirmed in Fig.~\ref{fig:plot_LT}.

For the intermediate case, $1<\alpha<2$, our type of single-file dynamics is
rather subtle.  As already shown in Ref.~\cite{lomholt}, despite the existence
of the scale $\langle\tau\rangle$ this regime behaves differently to the case
$\alpha>2$ 
\footnote{This behavior is in contrast to the single particle CTRW with
  $1<\alpha<2$ which has a simple Brownian MSD, just as for $\alpha>2$.}.
Our previous results in \cite{lomholt} would imply that $\langle n
\rangle\simeq t^{\alpha-1}$, suggesting the scaling ansatz
$H_n(t)\simeq t^{\alpha -1}f(n/t^{\alpha-1})$, where the prefactor is
again due to normalization \cite{REMM}. This approach would yield the
MSD $\langle x^2(t)\rangle\simeq t^{ \gamma(\alpha)}$ with
$\gamma(\alpha)=(\alpha-1)/2$. As shown in Fig.~\ref{fig:plot_MT}
(inset), this prediction for the scaling exponents does not agree well
with the simulations. An improved argument goes as follows: since the
random walk is unbiased, a given particle can equally well escape in
either direction from an interval confined by two blocking
particles. Thus this particle only needs to wait for the blocked
neighbor that moves first, corresponding to the minimum of two waiting
times drawn from the forward waiting time density $\psi_1( \tau)\simeq
\tau^{-\alpha}$. The distribution of this minimum time will have a
tail ${\tilde\psi}_1(\tau)=2\psi_1(\tau)\int_\tau^\infty
\psi_1(\tau')d\tau'\simeq \tau^{-2\alpha-1}$.\cite{schmittmann} The
resulting MSD for the labeled particle thus scales as
\begin{equation}
\label{eq:MSD_LT}
\langle x^2(t)\rangle\simeq t^{\gamma(\alpha)},
\end{equation}
where $\gamma(\alpha)=\alpha-1$ for $1<\alpha<3/2$ and $\gamma(\alpha)=1/2$
for $\alpha>3/2$. As seen from Fig.~\ref{fig:plot_MT}, this leads to an
improved agreement with the fitted exponents. We note that this argument would
become much more involved if we considered multiple escapes from blockage
events to further improve the agreement with the simulations data. This
argument using the minimum of two waiting times will not alter the MSD scaling
for $\alpha <1$, as the $\log t$ scaling was a result of the
aging of the waiting time distribution (i.e., its dependence on the time at
which the waiting began), a property that will be carried over in the
distribution of the minimum $\tilde{ \psi_1}$. The MSD scaling for $\alpha>2$ is
also unchanged by the modified argument above.

\begin{figure}
\includegraphics[width=8cm]{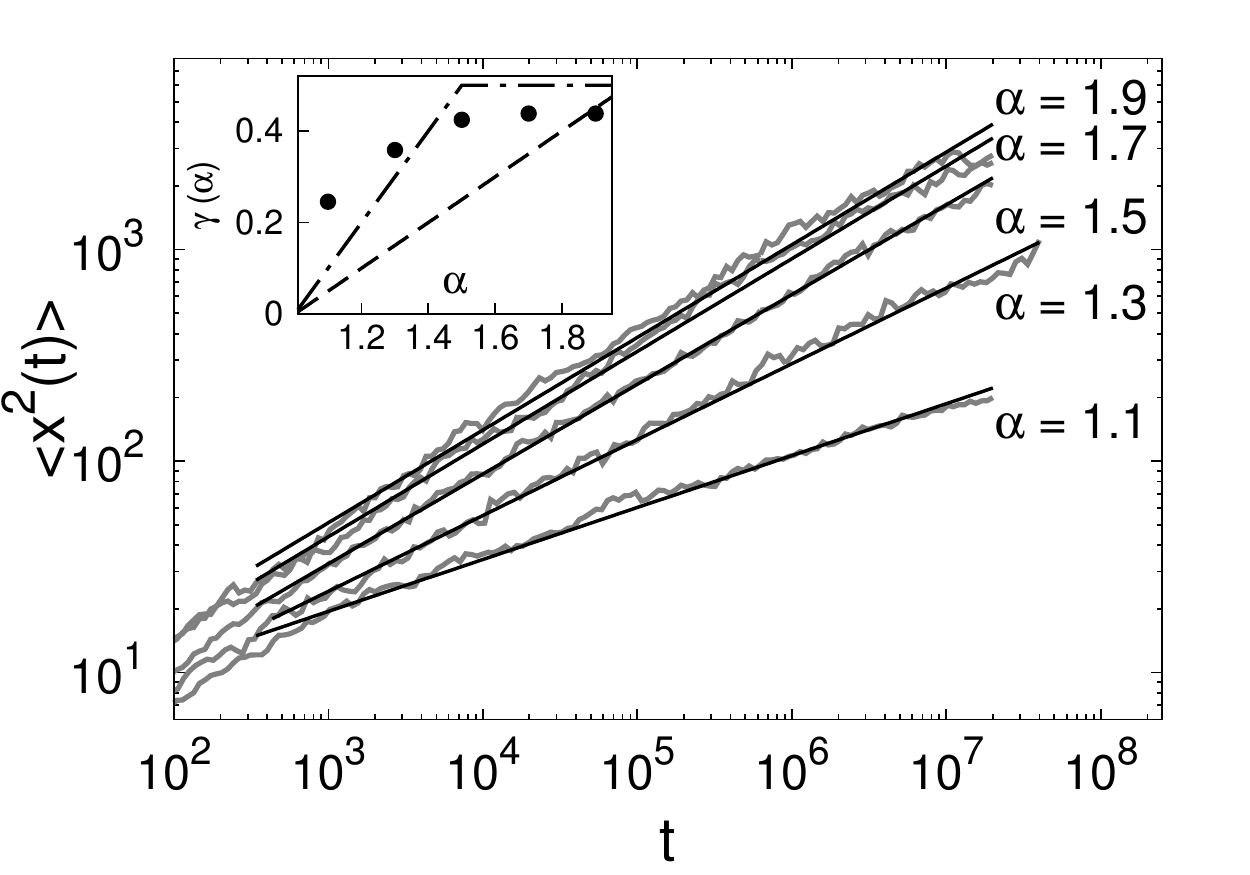}
\caption{MSD $\langle x^2(t)\rangle$ for a single file system with waiting time
distribution (\ref{wtd}) in the intermediate regime, $1<\alpha<2$. The fitted
scaling exponent $\gamma$ (inset) is compared to the predictions $\gamma(\alpha)
=(\alpha-1)/2$ (straight dashed line) and $\min\{\gamma(\alpha)=\alpha-1,0.5\}$
(kinked dash-dotted line) derived in the text. Other parameters are the same as in
Fig.~\ref{fig:plot_LT}.}
\label{fig:plot_MT} 
\end{figure}

\emph{Discussion.} We studied a physical model for the motion of
interacting (excluded volume) particles in an aging system. Building
on recent experiments of sticky particles moving along a
complementary, functionalized surface, we assume that each particle
performs a CTRW with a power-law waiting time distribution
$\psi(\tau)$. In particular, this scenario implies that each particle
carries an individual clock whose timer triggers motion attempts
according to this $\psi(\tau)$. As function of the laboratory time $t$
(master clock) we attempt to move the particle whose timer
expires first. Thus, while the update of the timers for each particle
is a renewal process, the excluded volume interactions lead to strong
correlations between the motion of the particles: when one particle
attempts to move and finds the neighboring lattice site occupied,
typically the blocking particle is caught in a long waiting time
period, and repeated attempts of motion by the mobile particle will be
required. In the long time limit, we demonstrated from scaling
arguments and extensive simulations that this many-body blockage
scenario leads to an ultraslow logarithmic time evolution of the MSD
of a labeled particle.

When the environment is less strongly disordered and the waiting time
exponent $\alpha>1$, the associated characteristic waiting time
$\langle\tau\rangle$ is finite. However, similar to biased CTRW
processes \cite{margolin}, there exists an intermediate regime for
$\alpha>1$, which still exhibits anomalous scaling: the MSD has a
power-law scaling with time, but the associated exponent is smaller
than the value 1/2 for Brownian (Harris) single file motion. Only when
the waiting time exponent $\alpha$ exceeds the value 2, the process
returns to Harris-type single file motion with $\langle x^2(t)\rangle\simeq
t^{1/2}$.

In Refs.~\cite{barkai,bandyopadhyay} another CTRW-based generalization of single
file motion was considered. However, their update rules for particles colliding
with a neighbor are very different. One way to view their process is that of a
castling, i.e., particles are allowed to move through each other (phantom
particles), while the labels of the particles switch in this castling. Thus
the labels will stay in the same order in the file and the tracer following a
specific label. Alternatively, the rule can be stated as particles switching
their clocks when they collide. Refs.~\cite{barkai,bandyopadhyay} found that
with this rule the generalized single file dynamics acquires the MSD $\langle
x^2(t)\rangle\simeq t^{\alpha/2}$ for $0<\alpha<1$. This result is fundamentally
different from our ultraslow result (\ref{eq:MSD_HT}), as we
explicitly consider excluded volume effects. We also mention that in
Ref.~\cite{ophir} a single file system of CTRW particles was considered, for
which clustering of particles and the asymptotic $\log^2(t)$ behavior of the MSD
were found. This approach is different from ours, in particular,
we do not observe any clustering.

Finally, we put the ultraslow time evolution discovered here in perspective to
other stochastic models with logarithmic growth of the MSD. The most famous
process is that of Sinai diffusion of a single particle in a quenched, random
force field in one dimension, leading to a $\log^4(t)$ scaling of the MSD
\cite{sinai}. In Sinai diffusion, deep traps exist at certain position of one
realization of the force field that cause the massive slow-down.  In our aging
single file system the strong interparticle correlations effect the
logarithmic time evolution. Logarithmically slow time evolution is also found
for a Markovian diffusion equation with exponential position-dependence of the
diffusion constant leading to a rapid depletion of the fast-diffusivity region
\cite{andrey}. The third class of stochastic systems with logarithmic time
evolution are renewal CTRWs with logarithmic waiting time distributions
\cite{julia}.

We expect our work to stimulate new research in the field of interacting
many-body systems in strongly disordered environments. 

TA and LL are grateful for funding from the Swedish Research
Council. RM acknowledges funding from the Academy of Finland (FiDiPro
scheme).


\begin{thebibliography}{99}
\bibitem{REMM} Note that for this intermediate case the spread grows
proportionally to the square root of the variance, i.e., there is no
deterministic relation between $n$ and $t$ as for the cases $0<\alpha<1$
and $\alpha>2$ \cite{lomholt}.
\end{thebibliography}
\end{document}